\documentclass[runningheads]{cl2emult}
\usepackage{epsfig}
\usepackage{graphicx}
\usepackage{cropmark}

\newcommand{\figPath}{.}

\newcommand{\ie}{{\it i.e. }}
\newcommand{\eg}{{\it e.g. }}
\newcommand{\Equ}[1]{(\ref{#1})}
\newcommand{\Fig}[1]{Fig.~\ref{#1}}

\begin{document}

\title*{Cooperation, adaptation and the emergence of leadership}
\titlerunning{Cooperation, adaptation and leadership}
\toctitle{Cooperation, adaptation and the emergence of leadership}

\author{Mart\'{\i}n G. Zimmermann\inst{1,2,}\thanks{zeta@df.uba.ar
and http://www.nld.df.uba.ar} 
\and 
V\'{\i}ctor M. Egu\'{\i}luz\inst{1,3,}\thanks{martinez@nbi.dk and
http://www.nbi.dk/CATS} 
\and 
Maxi San Miguel\inst{1,}\thanks{maxi@imedea.uib.es and
http://www.imedea.uib.es/PhysDept} 
}
\authorrunning{M. G. Zimmermann et al.}

\institute{Instituto Mediterr\'aneo de Estudios Avanzados
IMEDEA (CSIC-UIB), \\
Carretera de Valldemossa 7.5km, E-07071 Palma de Mallorca, Spain
\and
Depto. de Fisica--FCEN, Universidad de Buenos Aires,\\
Pabell\'on I Ciudad Universitaria,
1428 Buenos Aires, Argentina
\and
Center for Chaos and Turbulence Studies, The Niels Bohr
Institute,\\
Blegdamsvej 17,
DK2100 Copenhagen \O, Denmark}

\maketitle

\abstract{ A generic property of biological, social and economical
networks is  their ability to evolve in time, creating and
suppressing interactions. We approach this issue within the
framework of an adaptive network of agents playing a Prisoner's
Dilemma game, where each agent plays with its local neighbors,
collects an aggregate payoff and imitates the strategy of its best
neighbor. We allow the agents to adapt their local neighborhood
according to their satisfaction level and the strategy played.
We show that a steady state is reached, where the strategy and
network configurations remain stationary. While the fraction of
cooperative agents is high in these states, their average payoff
is lower than the one attained by the defectors. The system
self-organizes in such a way that the structure of links in the
network is quite inhomogeneous, revealing the occurrence of
cooperator ``leaders'' with a very high connectivity, which
guarantee that global cooperation can be sustained in the whole
network. Perturbing the leaders produces drastic changes of the
network, leading to {\em global dynamical cascades}. These
cascades induce a transient oscillation in the population of
agents between the nearly all-defectors state and the
all-cooperators outcome, before setting again in a state of high
global cooperation. }


\section{Introduction}

In the agent-based models used in Social Sciences, Economy and
Political Economy, agents interact directly with one another, and
a social macrostructure emerges from these interactions. The
implications of these models are easily studied with computer
simulations. In this computational approach it is very natural to
implement a network of interactions among the agents
\cite{Kirman99}. In addition,
such computer simulations permit to study the dynamical evolution
of the social structures.

In this context, an important question that is being addressed in
a number of ways is how the aggregate or global behavior {\em emerges}
from the individual characteristics of the agents.  A particular
aspect of this question is to understand if the global behavior is
determined by average commonly found agents or if a few individual
distinct agents can have a strong influence in the emerging
macrostructure. In the latter case such special agents play the
role of social leaders. Generally speaking, the influence of each
agent depends on the network of interactions with other agents, an
interaction being represented by a link between two agents.
These interactions can be restricted to a set of agents placed in
neighboring sites of a regular spatial lattice, can reach
arbitrary agents as in a random network or can occur through
intermediate ``small-world'' networks \cite{Watts98}. In most
cases, this network of interactions is fixed and given from the
outset. However, it is natural to consider situations in which the
network of interactions evolves dynamically adapting itself to the
emerging global structure.

In this paper we tackle the problem of how cooperation arises in a
dynamically evolving network of agents. The network adaptation
allows the emergence of an asymptotic state dominated by those
special agents which in the course of the dynamical evolution are
able to establish a much larger number of links than the average
agent.

The paradigm to study the emergence of cooperation has been the
Prisoner's Dilemma (PD) game.  Using evolutionary game theory
\cite{Weibull96}, it was shown \cite{Axelrod81,Axelrod84} that
cooperation may be sustained by a population of agents meeting
repeatedly through global random interactions. Two agents interact
playing the game and, according to their outcome, their strategies
are allowed to evolve. A second route to cooperative behavior,
pioneered by Nowak and May \footnote{\label{foot1}See for \eg
\cite{Nowak92,Lindgren94,Lindgren97,Huberman93,Nowak93}. A
cellular automata representation with several learning rules
is presented in Ref.~\cite{Kirchkamp95a}. For an extensive study
of the dynamics of a PD game with different strategies, evolution
of the strategies and different networks see Ref.~\cite{Cohen99}.}, comes
from the consideration of  {\em ``spatial games''}. In these
games every individual interacts only with a given subset from
the whole population  (e.g. the neighbors). The neighbors are
defined by a fixed network of interactions.  The spatial network
can promote the emergence of cooperation in situations in which
global non-cooperative behavior results if the interactions were
random and homogeneous. Here we also consider a spatial Prisoner's
Dilemma game. The novelty is that the group of agents with which
a given one interacts adapts {\em endogenously} during the
dynamical evolution. The adaptation of the network builds up
cooperation.

In our computer simulation of the game we implement two dynamical
rules. The first one is an action update: At each time step, each
agent plays the same strategy cooperate (C) or defect (D) with all
its local neighbors. Then the agents revise their individual
strategies and imitate the neighbors strategy with highest
aggregate payoff. Only a few agents will be found to be satisfied
and will keep their strategy. The second rule is the network
update:  Unsatisfied agents are allowed to change their local
neighborhood. Specifically, we let defectors break with a certain
probability any link with other defectors, and replace them with
new neighbors selected randomly from the whole network. The
motivation behind this rule is that two defectors playing a PD
game would certainly prefer to change its neighbor in order to
find a cooperator from whom to `exploit' a better payoff.

The results of our simulations show that the network of agents
reaches a steady state with a high degree of cooperative behavior.
The fraction of agents that cooperate depends only slightly on the
incentive to defect in the individual game. This behavior
contrasts with previous studies on ``spatial'' PD games where
partial cooperation was reached, but it was observed that the
fraction of agents which are cooperators strongly decreases as the
incentive to defect is increased. This feature results from the
adaptation of the network and it is reflected in the
non-homogeneous structure that it reaches during the dynamical
evolution. We will show that the process of ``searching''
cooperative neighbors performed by defectors, results in the
emergence of a leader agent, defined as the cooperator with the
largest number of links with other agents in the network.  When
the leader is {\em not} the wealthiest (i.e., the one with largest
payoff), the network is in an unstable situation and, depending on
the parameter $p$ measuring network adaptation, recurrent {\em
global cascades} may be observed. These cascades induce large
oscillations in the fraction of agents which are cooperators,
together with a large reorganization of the network. In most
cases, a final state with a high degree of cooperation is reached.
We have also tested the robustness of such cooperative state. We
find that a perturbation (spontaneous change of strategy) on a
non-leader usually results in a short transient dynamics returning
to the steady state. However, when a leader is perturbed, {\em
global cascades} may be observed in the system before a state with
a high degree of cooperation is recovered. This identifies the
importance of the highly-connected agents which play the role of
social leaders in the collective dynamics of the system.

The paper is organized as follows. The next Section defines the
spatial version of the PD game in an adaptive network. Section~3
describes our numerical results on the emergence of cooperation.
In Section~4 we discuss the evolution of the network and the
emergence of the "leader" agents. Finally in Section~5, we
summarize our results.

\section{Spatial Prisoner´s Dilemma in an adaptive network}

We consider the following framework: $N$ agents sit in the {\em
nodes} of an adaptive network $\Gamma$ where the {\em links}
define their neighborhoods. Each agent plays a PD game only with
those other agents directly connected by one {\em link}. In this
paper, we will restrict ourselves to the case of bidirectional or
undirected links, and interactions to first neighbors. Indirect
interactions with neighbors' neighbors have been studied for
example in \cite{Goyal99,Watts99,Jackson99}. Thus, two agents are
{\em neighbors} if they are directly connected by {\em one} link.
We define the {\em neighborhood} of agent $i$ as the subset of
$\Gamma$ which are neighbors of $i$, and we represent it as
neigh($i$); its cardinal is $K_i$. The {\em coordination number},
$K$, is defined as the average number of links per node
\begin{equation}
K=\frac{\sum_{i=1}^N K_i}{N}~. 
\end{equation} 
In this paper
we consider random networks $\Gamma$ with coordination number $K$
formed by distributing $K N/2$ links between pairs of nodes
$(i,j)$, with the constraint that $(i,j)=(j,i)$ (bidirectional
links). The resulting distribution of the number of links in the
network is Poissonian with the maximum located at the coordination
number $K$.

We denote by $s_i(t)=\{0,1\}$ the strategy of agent $i$ at time
step $t$, where $s_i=1$ corresponds to play cooperation (C), and
$s_i=0$ corresponds to defection (D), and will be referred to as
C-agents or D-agents, respectively.
The payoff matrix for a 2-agent PD game is shown in
Table~\ref{pdmat}, where it is standard to take
$b>\sigma>\delta>0$ and $b/2<\sigma$.

\begin{table}
\centering
\caption{Prisoners Dilemma payoff matrix}
\renewcommand{\arraystretch}{1.4}
\setlength\tabcolsep{5pt}
\begin{tabular}{c|c|c|}
  &         C       &      D \\
    \hline
C & $\sigma,\sigma$ & $0,b$ \\
  \hline
D & $b,0$      & $\delta,\delta$\\
  \hline\noalign{\smallskip}
\end{tabular}
\label{pdmat}
\end{table}

We consider the situation in which agents seek the largest
possible benefit from their local interactions in the network
$\Gamma$. We assume each agent plays the same strategy with all
its neighbors neigh($i$), and the strategy is updated by all the
agents at the same time; {\em synchronous} update. The time
evolution is as follows:
\begin{enumerate}
\item Each agent $i$ plays the PD game with each neighbor
using the {\em same} strategy $s_i$ and collecting a total
individual payoff $\Pi_i$,
\begin{equation}
\Pi_i = s_i \mu_i \sigma +(1-s_i)\left(\mu_i
b+(K_i-\mu_i)\delta\right)~,
\end{equation}
where $K_i$ is the number of links of agent $i$ and $\mu_i$ is the
number of neighbors of agent $i$ that are C-agents.

\item Agent $i$ revises its current strategy
at each iteration of the game  (\ie at  every time step), and
updates it by {\em imitating} the strategy of its neighbor with a
highest pay-off.  Agent $i$ is said to be {\em satisfied} if it
has the largest pay-off in his neighborhood. Otherwise it will be
{\em unsatisfied} and it will revise its strategy.

\item {\em \underline{Network Rule}}: each agent may adapt its
local neighborhood:

\noindent {\em if agent $i$ is an unsatisfied D-agent then with
probability $p$ breaks a link with a D-neighbor
$\alpha\in$neigh($i$), and replace it with a new agent $\beta$
uniformly from $\Gamma$.}

\end{enumerate}

This scheme leads to a time evolution of the structure of the
whole network, but the coordination number $K$ remains constant:
for each unsatisfied D-agent $i$, it will replace (on average)
$(K_i - \mu_i) p$ D-neighbors by new neighbors randomly chosen for
the whole set, and thus its local coordination number, $K_i$, will
not change; however, the replaced D-agents will lose one link and
the new selected ones will gain one link.

The network rule justification is based on the assumption that
given two D-agents playing a PD game, if they are unsatisfied,
then  they would prefer to exchange D-neighbors with the hope of
finding a new C-neighbor from whom to exploit a much better
individual payoff. This effectively amounts to ascribe a
``searching'' capability to D-agents. In our proposed setting the
searching is not optimized, in the sense that the searching is
random, so D-agents taking its chance to improve its payoff may
end up with a new C-neighbor with a larger payoff, forcing it  to
replicate this new strategy in the next time step.
In the same spirit, one could think that cooperators would also
have a preference to break links with other D-agents.  We prefer
to keep this asymmetry in roles so that D-agents may be described
as being competitive in nature, while C-agents remain
conservative\footnote{In \cite{Ashlock96} agents may refuse to play
with other agents irrespective of the strategy played.}.

The probability $p$ is  a measure of the adaptability of the
network. Dynamically, this parameter acts effectively as a time
scale for the evolution of the network. For $p=0$ the network does
not evolve and remains fixed, while for $0 \ne p \ll 1$ the
evolution is very slow as the adaptation takes over only after
several trials. For $p \sim 1$ the adaptation is done at the
same speed as the game is being played.   In other terms, $p$
represents a transaction cost composed of two parts: first, the
cost of breaking an  agreement and second, the cost of finding a
new partner and  that this new partner accepts the agreement.
One could separate these two costs, and would have a process of
breaking links (with a given probability $q$) and another
process of generation of links (with a probability $r$). However
for the sake of simplicity we consider these two processes as a
single one.

We have also investigated other variations of the network
adaptation rule. For example, instead of breaking links with any
of the unsatisfied D-neighbors, we also tested a rule which allows
an unsatisfied D-agent break with probability $p$ solely the link
with its D-neighbor with largest pay-off. The qualitative results
obtained with this adaptation rule are rather similar
\cite{Zimm00}.

It is worth making the following remarks:

\begin{itemize}
\item Links between satisfied agents do not change, which it seems to be a
reasonable assumption.

\item It is clear from the  network adaptation rule, that  not only D-agents
may {\em actively} change their neighborhood, but also C-agents will {\em
passively} evolve their own set of players by receiving new links from
``searching" D-agents.

\item In the present model, we do not take into account
spontaneous creation or destruction of links, therefore {\em the
total number of links in the network $\Gamma$ is conserved}.

\item In a standard 2-agent Prisoner's Dilemma game there exists a unique Nash
equilibrium (D,D). In a previous work \cite{Zimm00}, we considered
a variation of the PD payoff matrix with $\delta=0$, for
comparison with \cite{Nowak92,Lindgren94}. In this situation the
2-agent PD game has as pure Nash equilibria either (C,D), (D,C) or
(D,D). However the results in \cite{Nowak92,Lindgren94} indicate
that, at least for fixed regular network,  no qualitatively
difference is found in the spatial games when using $1 \gg \delta> 0$. 
\end{itemize}

In the next sections we present the results of computer
simulations of the model described above. We take as a free
parameter the incentive to defect $b$. We consider random initial
networks with coordination number $K=4$ and $K=8$ and several
values of the adaptability parameter $p$. We investigate, among
others, the following statistical measures: (i) the fraction of
cooperators, that it, the fraction of agents which are C-agents,
denoted by $f_C=(\sum_{i=1}^N s_i)/N$, (ii) the average payoff per
agent $\Pi=(\sum_{i=1}^N \Pi_i)/N$ of the whole network and the
distribution of payoff, (iii) the probability of having a link
between two C-agents, $p_{CC}$, between a C-agent and a D-agent,
$p_{CD}$, and between two D-agents, $p_{DD}$. These probabilities
satisfy:
\begin{equation}
1=p_{CC}+2p_{CD}+p_{DD}~. \label{def0} 
\end{equation}

Throughout this work, the parameter $b$ which controls  the
incentive to defect was varied in the range $1<b<2$, while the
other PD payoff matrix elements were fixed to  $\delta=0.1$ and
$\sigma=1$. Finally, if not otherwise stated, the network $\Gamma$
consists of  $N=10000$ agents, with an initial fraction of $0.6N$
C-agents randomly distributed in the network.

\begin{table}
  \caption{
Average fraction of C-agents,
$f_C$, for different initial random networks, with and without
network adaptation. The results are averaged over 10 different initial
conditions after $T=300$ time steps of evolution.}
\centering
\setlength\tabcolsep{10pt}
  \begin{tabular}{c|cc|cc}
  \hline\noalign{\smallskip}
         & $K=4$  & $K=4$  & $K=8$  & $K=8$  \\
    $b$ & $p=0$  & $p=1$ &  $p=0$ &  $p=1$ \\
  \noalign{\smallskip}\hline\noalign{\smallskip}
\renewcommand{\arraystretch}{1.4}
  1.05 & 0.89 & 0.942 & 0.95 & 0.994\\
  1.15 & 0.87 & 0.947 & 0.90 & 0.989\\
  1.35 & 0.59 & 0.920 & 0.58 & 0.988\\
  1.55 & 0.31 & 0.900 & 0.38 & 0.988\\
  1.75 & 0.09 & 0.885 & 0.03 & 0.983\\
  1.95 & 0.04 & 0.889 & 0.01 & 0.962\\
\hline\noalign{\smallskip}
\end{tabular}
\label{t1}
\end{table}

\section{Cooperation enhancement in an adaptive network}

The PD game with local interactions  in non-adaptive ($p=0$)
regular lattices, has been previously studied in
Refs.~\cite{Nowak92,Nowak93,Nowak94}. These studies showed how
partial cooperation can be sustained due to the local
interactions, in spite of memory-less strategies.  Several
extensions to this {\em spatial} model have been studied in the
literature. For example, introducing asynchronous updates
\cite{Huberman93} or introducing errors in the imitation process
\cite{Mukherji96}, the basic results persist \cite{Nowak94}.

For comparison with the adaptive network considered below, we give
in Table~\ref{t1} some numerical results for a fixed ($p=0$)
random network. Such network is our initial condition for the
adaptive network. We see that the average fraction of C-agents
decreases with an increasing value of the incentive to defect $b$.
Notice that depending on the characteristic coordination number
$K$, there is a critical $b=b^*$ such that for larger values of
$b$ the system reaches a state of all D-agents. That is, for
$b<b^*$ {\em partial cooperation} is supported in these networks.
We obtain for $K=8$ a critical  $b^*\approx 1.75$. The numerical
results also show that increasing the average size of the
neighborhood (average number of links per agent, $K$) the average
number of C-agents $f_C$ decreases faster with $b$.

\begin{figure}
\centering
\includegraphics[width=.90\textwidth]{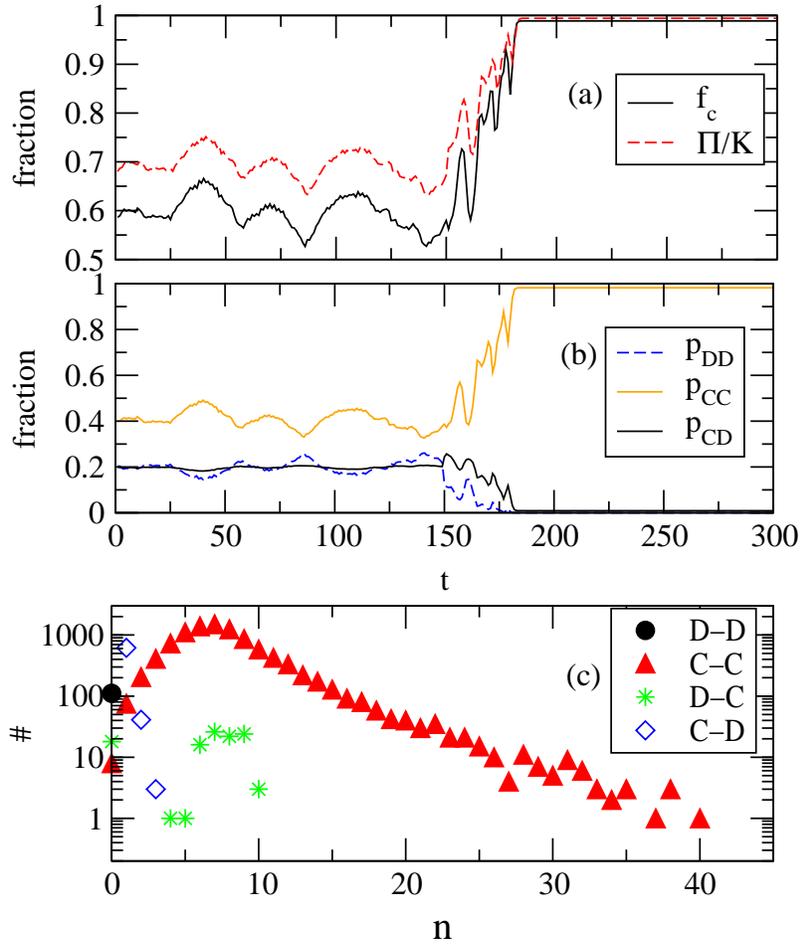}
\caption[]{(a) Time series of the fraction of cooperative agents
$f_C$ and average payoff $\Pi$ of the whole network. For $0<t<150$ the
 network is fixed $p=0$, while for $150<t<300$ the network
 is fully adapting $p=1$. (b) Corresponding time
series of the different links probabilities: two C-agents
($p_{CC}$), a C and a D-agent ($p_{CD}$), and two D-agents
($p_{DD}$) having a link. (c) Distribution of D-D-links,
C-C-links, D-C-links and C-D-links for the steady state.
($b=1.35$, $K=8$).} \label{ts}
\end{figure}


When the network is able to adapt with the outcome of the
individual games, either the dynamics settles onto a {\em steady
state} after some transient time, or the system reaches a full
defect state where all the agents are playing D and the network is
continuously evolving. To compare the asymptotic dynamics of the
fraction of cooperators between the adaptive and non-adaptive
case, we illustrate in Fig.~\ref{ts} a time series of $f_C$
evolving for $0<t<150$ in a fixed network ($p=0$), and then  for
$150<t<300$ with a fully adaptive network ($p=1$). We observe that
the fraction of cooperators increases as the adaptation is turned
on. Table~\ref{t1} illustrates this
behavior for other values of $b$. We conclude that the network
adaptation enhances a highly cooperative network. Also notice that
in the non-adaptive case, the fraction of cooperators fluctuates
slightly, while in the adaptive case a steady state is reached.

It is important to notice that in order to obtain the above
results, the {\em initial} fraction of cooperators in the network
has to be sufficiently large. In all our numeric computations we
took an initial fraction of $0.6 N$ of C-agents, which proved to
be a good number for the coordination numbers studied. This was
also noticed in previous spatial games and reflects the fact that
the cooperative strategy will replicate throughout the  network,
only if (the approximate) average payoff of C-agents is larger
than the one of D-agents. We will show below that in some
circumstances, perturbations of the network may destroy the
cooperative outcome and lead to a full defective network. Such
asymptotic state is a dynamical state, since the network is
continuously adapting but never finding a C-agent to exploit.

The steady state found in the adaptive network corresponds to a
stationary network structure and to a stationary configuration of
strategies. When the system reaches a steady state,
 there are necessarily no links between D-agents, thus $p_{DD} = 0$ (see \Fig{ts}(b)),
except in the unlikely case of having exactly the same aggregate
payoff. Also, in general we have that $p_{CD}\neq
0$. Therefore, the steady state is composed by a collection of
cooperators exploited by D-agents.

We  define a {\em chain of cooperators}, as a connected component
in $\cal N$ of sites occupied by cooperators in which each
cooperator except the last one is linked to a neighboring
cooperator with larger pay-off. All cooperators in the chain
except the last one
are actually unsatisfied, but as they imitate
the same strategy they were playing on the previous step, they
never change their relative payoff. When the system reaches a
steady state ($p_{DD}=0$), the only possibility for  D-agents is
to exploit the agents of a given chain and must necessarily be
satisfied. In terms of payoff we say that D-agents are ``passive''
local maxima, in the sense that they have the maximum payoff in
their neighborhood but nobody is imitating their strategy. In
summary, every cooperator chain should satisfy the following ordering in
terms of payoff:
\begin{equation}
\Pi_{l(i)}> \Pi_{r}\ge \Pi_i,
\qquad r \in \mbox{neigh($i$)}
\label{stable}
\end{equation}
where $i$ is a C-agent imitating from C-agent $l(i)$, and
$r$ is an exploiting D-agent of $i$. It is clear that the number
of different  steady states which one can construct satisfying
eq.~\Equ{stable} is huge for a fair number of agents $N$.

\begin{figure}
\centering
\includegraphics[width=.95\textwidth]{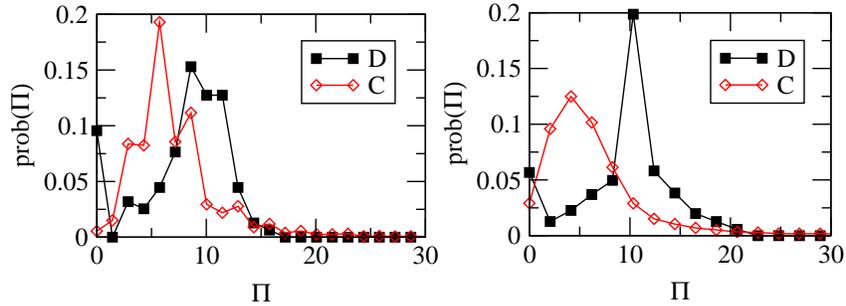}
\caption[]{Distribution of individual payoff for each
subpopulation (C and D-agents) in a final equilibrium state
(normalized  to each subpopulation). (a) $b=1.25$, (b)
$b=1.75$.  ($p=1.0$, $K=8$).}
\label{wealth}
\end{figure}

Finally, another salient feature of our cooperative network model
is the inhomogeneous distribution of payoff for each
subpopulation, which differs substantially  from the non-adaptive
network case. Figure \ref{wealth} displays such distributions
(normalized to the number of individuals in each subpopulation),
and reveals that although defectors are outnumbered by
cooperators, {\em on average they are wealthier}. This is an
interesting result, which indicates that, in the long run, the
searching capability of the D-agents rewards them. This behavior
is observed systematically in the parameter regime $1<b<2$ for
$K=8$.

The above result seems surprising from the point of view of
 the traditional replication dynamics
\cite{Weibull96} used in evolutionary game theory, because one
could conclude from  \Fig{wealth} that D-agents should dominate
the whole population.  But our results indicate that the final
highly cooperative state is not determined by average agents in
the system, but rather by a small subset of those maximally
connected C-agents. In the next section we discuss the important
role played by these {\em cooperative leaders}, defined as those
C-agents with the largest number of links.

\section{Dynamics of network evolution and emergence of leaders}

The network rule allows for the evolution of the connectivity of
every agent and permits that the network reaches a steady state.
The distribution of links in this steady state displays the
heterogeneous structure of the network. Figure~\ref{ts}(c) shows
the distribution of links between two C-, D-, and between D-C and
C-D agents. Notice the broad band distribution of links between
two C-agents, which may reach as 5 times the average connectivity
of the network. Defectors, on the other hand, are shown to be
connected only to C-agents, and  have a narrow distribution
centered at $K$.

The tail of the distributions of links identifies a very small
number of C-agents with a large number of links to other C-agents
We define the C-agent with the maximum number of links as {\em the
leader}. In a steady state this agent should be satisfied and
leading a chain of cooperators. The number of C-agents which can
fit on a given chain connected to the leader may be huge.

\begin{table}
  \caption{
 Maximum number of links of the leader agent ($K_\alpha$), the
 D-agent with a largest number of links
 ($K_\beta$), and corresponding payoffs ($\Pi_\alpha$, $\Pi_\beta$) for different values
of $b$ in a steady state.  The results are averaged over 30
different initial conditions, $K=8$ and $p=1$.} 
\centering
\setlength\tabcolsep{10pt}
  \begin{tabular}{ccccc}
  \hline\noalign{\smallskip}
  $b$  & $K_\alpha$  & $K_\beta$ & $\Pi_\alpha$  & $\Pi_\beta$     \\
  \noalign{\smallskip}\hline\noalign{\smallskip}
\renewcommand{\arraystretch}{1.4}
  1.15 & 25.8 & 11.2 & 25.8 & 12.9  \\
  1.35 & 40.5 & 11.7 & 40.5 & 15.8  \\
  1.55 & 54.1 & 11.8 & 54.1 & 18.3  \\
  1.75 & 56.9 & 13.0 & 56.9 & 22.8  \\
  1.95 & 72.1 & 14.0 & 72.1 & 27.4  \\
\hline\noalign{\smallskip}
\end{tabular}
\label{t2}
\end{table}

We have calculated the number of links of the leader agent for
different values of the incentive to defect $b$ in a steady state.
Table~\ref{t2} gives a measure of the connectivity of the leader,
labeled $\alpha$, compared with the number of links of the most
connected D-agent labeled $\beta$ (recall that all connections are
exploiting other C-agents), with their corresponding payoffs.
Notice that for increasing $b$ the number of links of the leader
increases while it remains essentially constant for the defector
with largest number of links.

The leader agent leads the cooperative collective state of the
system in several ways. On one hand, the leader favors the
creation of chains of cooperators. On the other hand, and provided
there are links between D-agents, leaders are selected through the
dynamics of the adapting network, and are a direct consequence of
the ``searching'' done by D-agents. Consider for example a D-agent
$j$ which exchanges one of its D-neighbors with the leader
$\alpha$. Assume that $\Pi_\alpha>\Pi_j$. In the next time step
the D-agent will become a cooperator by imitation and
\begin{equation}
K_\alpha(t+1)=K_\alpha(t)+1 \label{cres} 
\end{equation} 
Whenever the payoff
of the leader is the largest payoff in the whole network its
connectivity increases as described by \Equ{cres}.

However, {\em an unstable situation occurs whenever the leader
does not have the largest payoff in the whole network}. If the
leader $\alpha$ receives a D-neighbor with a {\em larger} payoff,
in the subsequent time steps a drastic event happens, for $alpha$
and its associated chain will imitate the D-strategy. If after
this cascading imitation there are C-agents left in the network, a
new leader with a fewer number of links will be selected;
otherwise a full defective network may be reached.  This indicates
the sensibility of the network structure to small perturbations on
individual specially well-connected agents: A local event
associated with a particular individual propagates in
macrodynamical avalanches into the full network.

The phenomenon described above does not exclude the possibility
that a D-agent $j$ selects a C-agent which would satisfy
\Equ{stable}, allowing agent $j$ to become a passive local maximum
and exploit the C-agent. In fact, it is by this mechanism that
D-agents also increase their payoff.

\begin{figure}
\centering
\includegraphics[width=.95\textwidth]{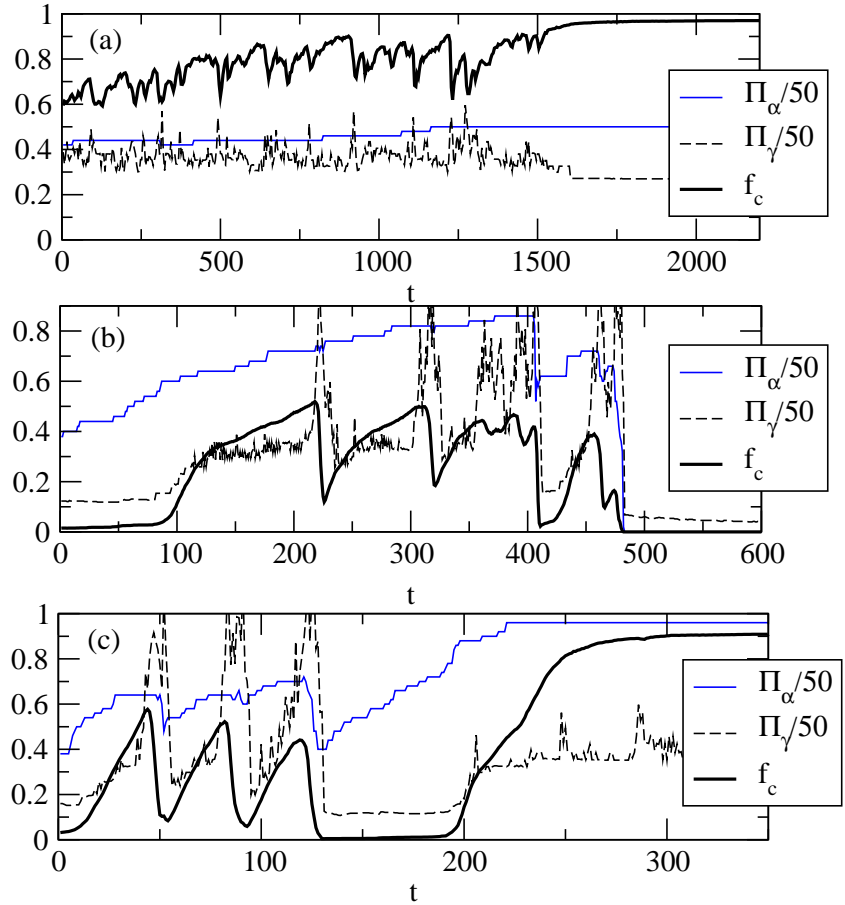}
\caption[]{Time series of $f_C$ and a re-scaled payoff of the
cooperator leader ($\Pi_\alpha$) and the D-agent with maximum
payoff ($\Pi_\gamma$). (a) $p=0.005$, $b=1.35$, (b) $p=0.01$,
$b=1.75$, (c) $p=0.05$, $b=1.75$ ($K=8$).} \label{posta4}
\end{figure}

We have performed several computer simulations to visualize the
dynamic evolution of the network which we have just discussed.
Each panel of \Fig{posta4} shows a comparison of the payoff of the
leader $\alpha$ with the payoff of the D-agent with largest payoff
(labeled $\gamma$\footnote{We remark that at every time step a
different D-agent may become the agent $\gamma$ with largest
pay-off. This is a consequence of the competitive nature of
D-agents in contrast to the conservative nature of C-agents.}),
together with the evolution of the fraction of cooperators in the
network.

For small values of the adaptability parameter $p$, as shown in
panel (a), the typical time for the network to reach a steady
state is very large, and the  leader $\alpha$ increase its
connectivity via \Equ{cres} very slowly. The fraction of
cooperators increases on average as the leader increases its
number of links. On the other hand for $p\approx 1$, the typical
time to reach a steady state is very short (see \Fig{ts}), and in
general the leader is always the wealthiest agent of the whole
population.

 An interesting intermediate regime occurs for $p\approx 0.05$
and high incentive to defect $b$, where a competition between the
cooperative leader and the most wealthy defector arises. In this
regime the number of links of the leader does not grow so fast and
defectors may get a significant large payoff due to the high value
of $b$. Panel (b) and (c) illustrates this situation. It is seen
that there are time intervals in which the leader is not wealthier
than the D-agent $\gamma$ with largest payoff. Whenever this event
occurs, large global cascades involving a large fraction of the
whole population are observed, with a significant fraction of the
whole population being affected. The initial dropout of the
fraction of cooperators affects immediately the payoff of
exploiting D-agents and $f_C$ reaches a minimum approximately when
the leader $\alpha$ becomes once again the wealthiest agent in the
network. Once this stable situation is re-established, the leader
may again increase its number of links until all D-agents have
links exclusively with other C-agents ($p_{DD}=0$).

An extreme example is observed in panel (c) at $t\approx 150$,
where the network is composed of mostly D-agents together with a
very wealthy leader. The leader is able to increases its number of
links  by the intense (unsuccessful) ``searching'' done by
D-agents, and by  $t\approx 200$ the fraction of cooperators also
increases. This recovery shows the importance of a wealthy leader,
which enables a full cooperative final outcome. Another situation
worth mentioning is that whenever strong competition between the
leader and the wealthiest D-agent occurs, there is a possibility
that the system ends in a full defective network. This is what is
shown happens in panel (b) of  \Fig{posta4}. We remark that
provided the initial fraction of C-agents is large, the full
defective network is rarely reached for high enough $p$.

\begin{figure}   
\centering  
\includegraphics[width=.95\textwidth]{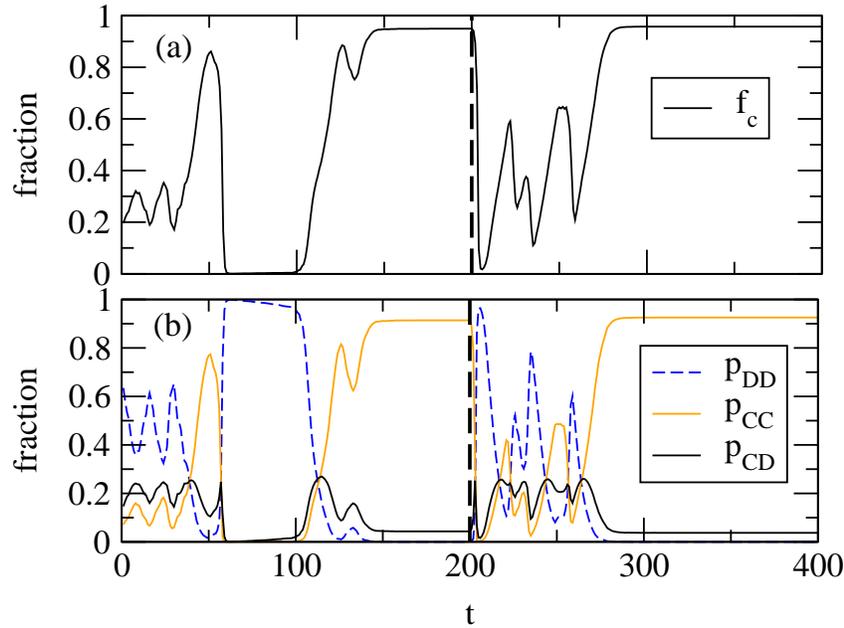} 
\caption[]{
Time series of $f_C$, where at $t=200$ the leader agent changes
strategy from C to D. Parameter values: $b=1.75$, $K=8$,
$p=0.05$.} \label{switch}
\end{figure}

The above results show that the dynamic evolution of the network is
intimately related to the fate of the leader.  Another possible
test, is to study how {\em noise} affects the network dynamics. If
the noise is in the form of selecting a random agent and
spontaneously changing its strategy, then the dominance of
C-leaders is found to remain for nearly the whole range of
$1<b<2$, for a sufficiently small noise intensity. However when
the probability of a spontaneous change of strategy is increased,
we find that a transition to the full defective network becomes more
probable\footnote{In fact a full
defective network is reached for $b>b^{**}$, where the critical $b^{**}$
depends on the noise intensity.}. Clearly the leader may suffer such
perturbations, and induce large cascades in the system. If such
drastic perturbations have a small probability, the system has
time to reach the cooperative outcome before the new leader is
knocked down again. The simulation described in \Fig{switch}
illustrates how the system reaches a steady state at $t\approx
180$, and a spontaneous change of strategy was applied to the
leader: large cascades results and a final cooperative outcome is
again recovered. The transient oscillations observed in
\Fig{switch} at $t\approx 50$ before a steady state is reached,
as well as the ones originated by the change of strategy of the
leader constitute a Sysiphus effect. The drop in the fraction of
cooperators comes together with a large increase of $p_{DD}$.
Thanks to its adaptability, the network reacts creating more C-C
links. However the attempts to build up a large cooperative
behavior are not always successful and the system goes through
oscillations in which non-cooperative behavior with large values of
$p_{DD}$ is recovered. The frustrated attempts to build
cooperation indicate that for cooperation to be robust, it has be
built upon a specific networks of links. In the frustrated
attempts to reach a collective stable cooperative state, the
fraction of cooperators becomes large, but the spatial arrangement
of links in the network is not the proper one.

\section{Discussion}

We have introduced a model of cooperation on an adaptive network,
where cooperation is highly enhanced with respect to the situation
of a fixed network. The network adaptation involves exclusively
the D-agents, which in some sense are allowed to ``search" for new
neighbors, in the hope of finding C-agents to be exploited. Our
study shows that this mechanism leads to a global cooperative
state of the network. The asymptotic state reached by the system
is a steady state in which the network structure and the average
payoff per agent $\Pi$ remains stationary. However, most agents
are unsatisfied, and continuously imitate the strategy of their
neighbors with highest payoff (most of them C-agents). The
structure of the network can be understood in terms of chains of
cooperators with D-agents exploiting some cooperators.

The network adaptation that we have implemented in our spatial
game gives rise to the emergence of a leader, a cooperator with
maximum number of connections. Whenever the leader is the
wealthiest agent in the network, and defectors are still
``searching'', the leader may increase its number of links with
other cooperators. However, if some defector becomes wealthier
than the leader, an unstable situation occurs and large global
cascades may take place. Such cascades can also be induced if 
``noise'' in the form of 
spontaneous change of the strategy is allowed.

We finally note that the dynamics described above cannot be
explained in terms of average agents. In the final mostly
cooperative state the average wealth of defectors is larger than
the average cooperators wealth. The final collective state is
dominated by rare individuals which build-up cooperation in the
whole population because they have a number of links which is far
from the average.

\section*{Acknowledgments}
 MGZ acknowledges financial support from FOMEC-UBA.
 VME acknowledges financial support from the Danish Natural
Research Council. MSM acknowledges financial support from the
Spanish Ministerio de Ciencia y Tecnolog{\'\i}a project
BFM2000-1108.


\end{document}